\documentclass{elsart}

\usepackage{natbib}

\usepackage{graphics}
\usepackage[a4paper]{hyperref}
\usepackage{graphicx}
\usepackage{epsfig}

\usepackage{amssymb}

\begin{document}

\begin{frontmatter}


\title{A numerical code to study the variability of Blazar emission}
\author{A. Tramacere},
\ead{andrea.tramacere@fisica.unipg.it}
\author{G. Tosti}
\ead{gino.tosti@fisica.unipg.it}
\address{Physics Department and Astronomical Observatory, University of Perugia, Via Pascoli I-06100 Perugia, Italy}

\begin{abstract}
We present a numerical code, written in C, which can be used to simulate or to analyze the emission of \emph{Blazars} over the entire electromagnetic spectrum. Our code can reproduce: Synchrotron emission, Inverse Compton emission (IC) (Thomson \& Klein-Nishina regime), External Compton emission (EC), accretion disk variability using  a Cellular Automata (CA) algorithm, temporal evolution of the emitting plasma energy distribution, flaring phenomena, light curves in the rest and in the observer frame (taking account for time crossing effects).
\end{abstract}
\end{frontmatter}
\section{Introduction}
\emph{Blazars}, are objects which emit energy in the entire electromagnetic spectrum, and are characterized by a variability which scales from long to short period. To understand the real nature of \emph{Blazar}, we need multiwavelenght simultaneous data, and we need to be able to analyze these data and to do numerical predictions to discriminate between the theoretical models. To the light of these simple but constraining remarks, we decide to write a simulator able to simulate the \emph{Blazar} energetic and temporal behavior. A large effort has been produced to ensure a numerical precision enough high to use this code to analyze data too. This work represents just the first step of our project, which will have many future developments. To make our simulation as complete as possible, we decided to implement the simulation of disk variability too. In the section \emph{2} we present the simulation of the disk, section \emph{3} is dedicated to show how we can reproduce light curve both for SSC and EC model, taking into account for time crossing effects.

\section{Simulation of a light curve of an accretion disk, by cellular automata }
\begin{figure}  
\begin{center}
\includegraphics[width=14cm]{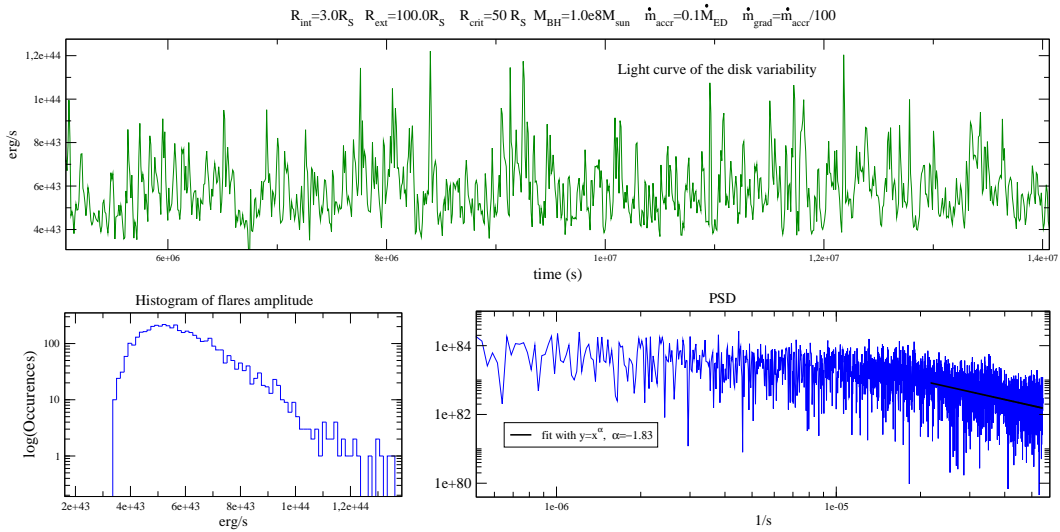}
\caption{Simulation of the disk variability using CA.}
\label{fig:disk}
\end{center}
\end{figure}

The rapid quasi-fractal variability of \emph{Blazar} is one of the main signature for their temporal behavior. This kind of variability is common to many astrophysical objects (sun spot, galactic accretion disk around black hole ecc....).   The Cellar Automata (CA) \citep{Mineshige}, \citep{Takeuki} is a fast and efficient way to simulate this variability, without the need of solving differential equations. In this work we used CA to simulate the variability of the accretion disk emission. In \emph{Blazar} the jet axis is close to the observer line of sight, so luminous fluxes are dominated by the jet emission. This led us to the understanding  that the  X-ray fluctuation, are due mostly to the emitting blob, and not to the disk. Therefore to have a quantitative indication about the behavior of an accretion disk acting around a super massive Black Hole (BH) ($m>10^8M_{\odot}$), we try to simulate the disk emission of the Seyfert galaxies, which have not significative contribute from jet emission. The analysis of the X-ray light curves for these objects \citep{Hayashida}, shows a behavior similar to the \emph{1/f} noise, so the PSD of these light curves has almost the following form $PSD \propto f^{-\alpha} $ with  the slope $\alpha$ in the range $1-2$. We than model our disk, tuning the input parameter, to obtain all the physical characteristic that are supposed to belong to an accretion disk around a super massive BH. We performed many simulation, and in fig \ref{fig:disk} we show just the result closest to the condition we believe more realistic. To model the surface density of the disk, we used  a parametrization \citep{Takeuki1}, which relates the steady \citep{Shakura} with the critical regime in the following way
\begin{equation}
\Sigma_{crit}(r)=\Sigma_{steady}(r)\cdot[1+\eta(r-R_{crit})/(r_{out}-r_{inner})]
\end{equation}
The parameter $\eta$ is fixed at the value of 0.5, while the value of $R_{crit}$,  has been varied from 10 to 90 $R_S$. We found that the slope of the PSD, varies with $R_{crit}$. The value of the slope closest to those of the Syfert galaxies is for $R_{crit}=50$. We noted how the slope change with the gradual mass flow  $\dot{m}_{grad}$ too, in good agreement with other simulations \citep{Xiong}.

\section{Reprocessing the disk variability by the blob, a mechanism to explain ERC variability}
\begin{figure}  
\begin{center}
\includegraphics[height=6cm]{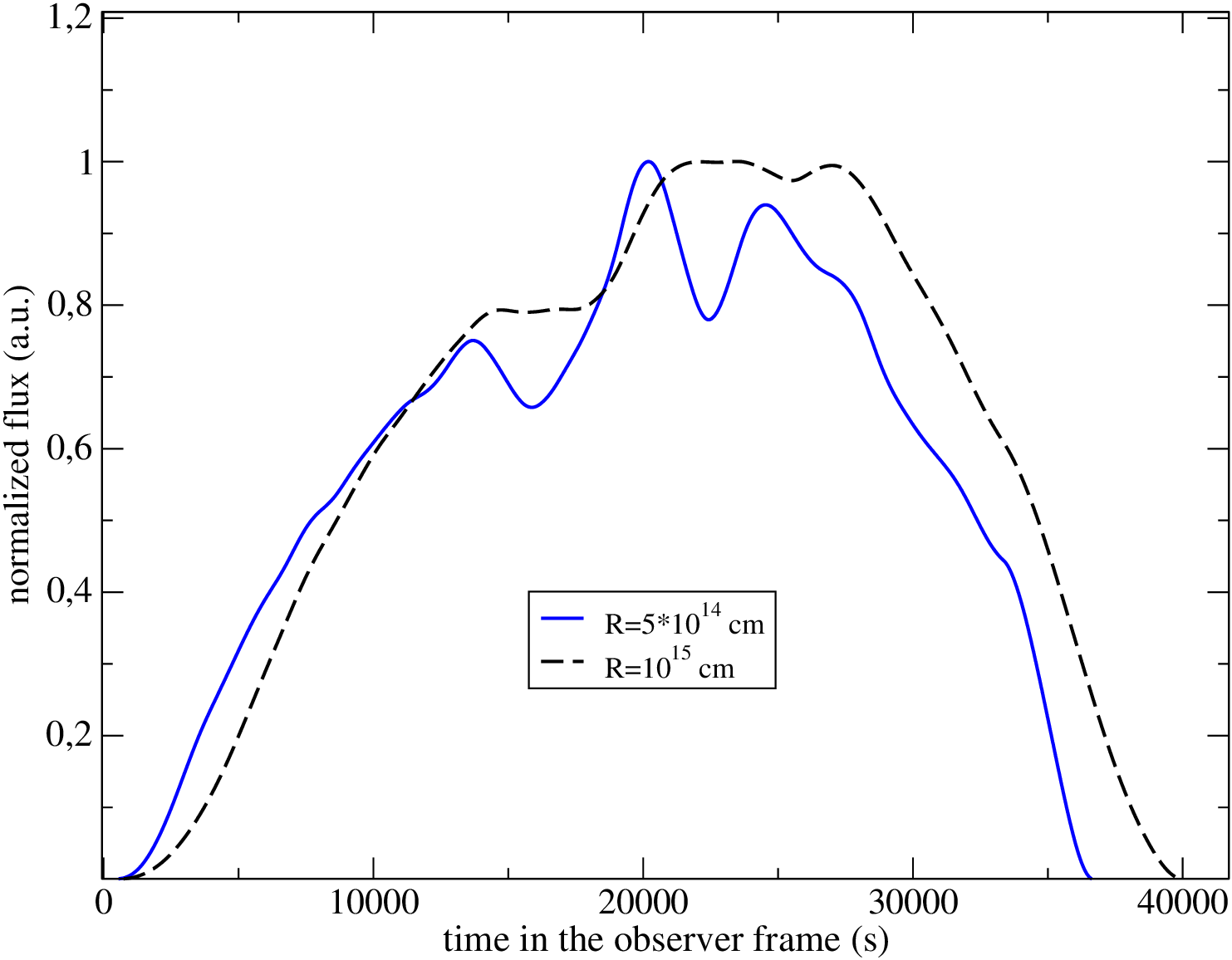}
\caption{Simulation of EC light curves taking into account for time crossing effects }                                         
\label{fig:extc}
\end{center}
\end{figure}

We simulate the temporal evolution of the emitting plasma in the blob \citep{Kataoka},\citep{Chiaberge}, solving numerically the Fokker Planck equation, following the method of Chang \& Cooper \citep{Chang}, and obtain light curves both for synchrotron \citep{Inoue}, \citep{Rybicky} and IC/EC \citep{Band}, \citep{Blumenthal}, \citep{Deremer}  emission. Than we reprocess these light curves taking into account for time crossing effects. Here we present a simulation in which the light curve of the disk (showed in  fig. \ref{fig:disk}) constitutes the external photon field for the EC mechanism. We than assume that the Broad Line Region (BLR) reflection of the disk luminosity (with an reflectivity $\chi$=0.0001) generates a quasi-isotropic photon field, and do not take into account  for the radiation coming straightly from the disk. This light curve has a typical time scale in the disk frame of about $10^4s$. Our simulator take into account for the relativistic transformation for time and frequencies, between the disk-BLR frame and the blob frame and between the blob and the earth. In this simulation we assume that the blob  moves toward the BLR with a typical $\Gamma=10$, so we use a beaming factor $\delta=10$ for every the transformation. In fig \ref{fig:extc} we report the result for two different values of $R$: $10^{15}cm$ and $5\cdot10^{14}cm$. The number of slices into which we divide the blob is such that the time crossing of each slice in the blob frame  is of 100 s, that is much shorter than $10^4/\delta s$ which is the time scale variability of the disk seen by the blob frame. 

\section{Conclusions}
One of the main goal of our work, was understanding the connections between the accretion disk variability, and it's reprocessing by the emitting blob. We observed that the luminosity fluctuations, that are present in the disk, are seen by the observer, if the blob radius is such that photon time crossing is not much larger than typical disk variability time scale. In the future we aim to explore the connection between the disk instability and the disk fueling to the light of the internal shock scenario, to simulate the synchrotron rapid variability  too, using a self consistent physical model.




\begin{thebibliography}{}
\bibitem[Inoue \& Takahara (1996)]{Inoue} Inoue, S.,  Takahara F. 1996 ApJ, 463,555
\bibitem[Band \& Grindlay (1985)]{Band} Band, D. L., \& Grindlay, J. E. 1985, ApJ 298,128
\bibitem[Blumenthal\& Gould (1970)]{Blumenthal} Blumenthal, G. R.,\& Gould, R. J., 1970, Rev. Mod. Phys 42,237	
\bibitem[Dermer \& Schlickeiser (2002)]{Deremer} Dermer C. D., \& Schlickeiser, R.,astroph/0202280       	
\bibitem[Rybicki \& Lightman (1979)]{Rybicky} Rybicki, G. B., \& Lightman, A., P.,1979 Radiative Processes in Astrophysics, Jhon Wiley \& Sons
\bibitem[Kataoka, J., et al. (2000)]{Kataoka}Kataoka, J., et al. 2000, ApJ 528,243	         
\bibitem[Chang \& Cooper (1970)]{Chang}Chang, J. S., \& Cooper, G. 1970, Journal of Computational Physics, 6, 1
\bibitem[Chiaberge \& Ghisellini (1999)]{Chiaberge}Chiaberge, M., \& Ghisellini, M. 1999, MNRAS 360,551
\bibitem[Mineshige, Ouchi \& Nishimori (1994)]{Mineshige}Mineshige S., Ouchi, B. N., \& Nishimori H. 1994 ,PASJ 46,97
\bibitem[Takeuchi, Mineshige \& Negoro (1995)]{Takeuki}Takeuchi, M., Mineshige, S.,\& Negoro, H. 1995, PASJ 47,617 
\bibitem[Takeuchi \& Mineshige (1997)]{Takeuki1}Takeuchi, M., \& Mineshige, S., 1997, ApJ 486, 160
\bibitem[Shakura \& Sunyaev (1973)]{Shakura}Shakura, N. I., \& Sunyaev, R., A. 1973,  A\&A  24,337
\bibitem[Negoro et al. (1995)]{Negoro}Negoro, H., et al. 1995, ApJ 452, L49
\bibitem[Xiong,Witta \& Bao (2000)]{Xiong}Xiong, Y., Witta P. J.,\& Bao, G., astroph/0009424
\bibitem[Hayashida et al. (1998)]{Hayashida}Hayashida, K., et al. 1998, ApJ 500,642-649
\bibitem[Tramacere A., (2003)]{Tramacere}Tramacere A., 2003, in progress		
\end{thebibliography}
\end{document}